\documentclass[preprint,showpacs,amsmath,amssymb]{revtex4}
\usepackage{graphicx}
\usepackage{amssymb}
\newcommand{\etab}{\eta_{\mathrm{bulk}}}
\newcommand{\etad}{\eta_{\mathrm{drop}}}
\newcommand{\etaf}{\eta_{\mathrm{eff}}}
\begin{document}

\title{A convective instability mechanism for quasistatic crack branching in a hydrogel}
\author{T. Baumberger and O. Ronsin}
\affiliation{INSP, UPMC Univ Paris 06, CNRS UMR 7588, 140 rue de Lourmel, 75015 Paris France}

\date{\today}
\begin{abstract}
Experiments on quasistatic crack propagation in gelatin hydrogels reveal a new
branching instability triggered by wetting the tip opening with a
drop of aqueous solvent less viscous than the bulk one.  We show that
the emergence of unstable branches results from a balance between the
rate of secondary crack growth and the rate of advection away from
a non-linear elastic region of size $\mathcal G/E$ where $\mathcal G$
is the fracture energy and $E$ the small strain Young modulus. We
build a minimal,
predictive model that combines
mechanical characteristics of this mesoscopic region and  physical features of the process
zone.
It accounts for the details of the stability diagram and lends
support to the idea
that non-linear elasticity plays a critical role in crack front
instabilities.
\end{abstract}
\pacs{46.50.+a,62.20.mt}
\maketitle

\section*{Introduction}

Recent developments in tissue engineering \cite{Tissue} have raised polymer
hydrogels to the status of  genuine, {\it structural} materials,
suitable for load bearing
applications, such as scaffolds for in vivo tissue regeneration.
Disregarding the biological issue, an essential step towards the rational
design of such hydrogel-based systems is to understand how these ``soft"
solids break. This means not only  relating
their micro-structural features to their mechanical
strength, but also describing the way failure will eventually
proceeds, i.e. how damage will pervade a sample set in a given
mechanical as well as physico-chemical
environment.

Hydrogels
are network-based, solvent-swollen materials. The covalent (resp.
non covalent) nature of the
cross-linking bonds plays an essential role in chemical (resp.
physical)  gel strength. Chemical gels require C-C backbone
scission to break and exhibit quasi rate-independent energy dissipation
during crack propagation. Hence, frac\-ture usually proceeds
dynamically, i.e. accelerates until reaching values in the vicinity
of the Rayleigh wave velocity of the material. Chemical gels are
therefore termed brittle and as such have received much attention
recently for their ability to mimics the dynamic fracture
pheno\-menology of hard brittle solids, albeit at much lower
velocities \cite{Cohen}.
This is in
contradistinction with physical gels in which weaker crosslinks act
as mechanical fuses preventing chain scission. Disrupting
large crosslink structures \cite{Triblock} and/or pulling chains out of the gel
matrix \cite{Model} are usually strongly dissipative,
rate-depen\-dent breaking processes which are responsible for the amazing
resistance to fracture of these materials, mostly consisting of
solvent with only a small fraction
of   polymer. Steady, quasistatic crack propagation is readily achieved
with physical gels. This makes them suitable for detailed
investigation aimed at unraveling the physical, dissipative mechanisms at work.

Notwithstanding these fundamental differences,
both physical and chemical gels have in common their softness i.e. their
large elastic compliance. Consequently, the fracture process
involves finite crack tip opening displacements and large
deformations over a wide region ahead of the tip. This situation is
markedly at odds with the requirement of linear elastic fracture mechanics
(LEFM) that small strain, {\it linear} elasticity prevails
everywhere apart from  a near tip ``process zone"
where dissipative bond-breaking
occurs.
Owing to the powerful ability  of LEFM to predict the onset of
failure with minimal material-dependent input \cite{Lawn}, though,
the role played by the near crack-tip, non-linear elastic (NLE) zone
where LEFM breaks down,
has been long overlooked.
This, nevertheless, cannot be ignored any more when dealing with two important
issues:

{\it Elastic crack blunting}  --- In soft elastic solids which
can sustain  bond-breaking
stresses much larger than their small-strain  Young modulus $E$,
crack tips tend to blunt, i.e. to develop large radii due to
purely elastic deformations  \cite{Blunting}, thereby mitigating
remote stress intensification.
A strict LEFM approach  would lead to the paradoxical conclusion
that stresses, which are predicted to  plateau at values of order
$E$,  remain too low
to initiate rupture \cite{Blunting}. Strain-hardening, a NLE characteristics shared by
strongly stretched polymer networks, has been proposed as a likely
way of restoring stress concentration in the near tip region.

{\it Crack front instabilities} --- Smooth crack propagation, either
quasistatic or dynamic,  is rather the exception than the rule
in fracture of soft elastic solids. Branching \cite{Cohen}, splitting
\cite{Triblock,Morpho}
or oscillating \cite{BenDavid,MarderRubber}
cracks are commonly reported phenomena which LEFM fails to explain \cite{Marder}.
Again,  the existence of the NLE zone, introducing a new length scale
in the fracture problem, has
been invoked as the missing
ingredient for predicting the onset
of a front instability and its characteristic features
\cite{Livne,HuiTip}. It is worth noting that crack tip blunting and splitting
have already been evoked by Gent
\cite{Gent} as possible {\it causes}  of the high tear
strength of visco-elastic elastomers (unswollen chemical gels). He
listed them amongst several unresolved issues in rubber fracture.
Indeed, despite recent advances in numerical and theoretical description
of the NLE zone, elastic blunting  and front instabilities remain
widely open problems which probably transcend the case of
hydrogels.

These materials, which feature no noticeable linear visco-elasticity over
a wide frequency range, are not expected to  exhibit  bulk energy dissipation
during crack propagation but ra\-ther  localized dissipation in
a near tip process zone of extension $d$. In the case of gelatin
gels, it has been argued, based on experiments \cite{EPJE} that $d\sim 100$ nm. The
size of the  NLE
zone  scales with the
natural length in fracture problems $\mathcal L = \mathcal G/E$ where
$\mathcal G$ is the energy release rate (free energy
released by the advance of a unit area of crack) which, for a
quasistatic crack,  identifies
itself with
the dissipated fracture energy. In the case of
gels, which are ``tough" solids with relatively large $\mathcal G$
and low $E$, $\mathcal L$ ranges typically between $100\,\mu$m and a
few mm (as compared, e.g. to a few \AA~  for brittle silica glass).
Thus, fracture in hydrogels exhibit a clear hierarchy of relevant
length scales between the microscopic $d$, the mesoscopic $\mathcal L$
and the macroscopic system size $h$~:

$$d\ll \mathcal L\ll h$$
This configuration, which extends the so called ``small scale yielding"
hypothesis \cite{Broberg} to the NLE case, plays a central role in
the following.

This article aims at presenting experimental
evidence of a previously undescribed  branching instability in a
physical gelatin gel. The clear separation of length scales enables us to
unravel NLE effects from dissipative mechanisms. The
instability is therefore
amenable to a physical interpretation which accounts --- though
schematically --- for the elastic blunting of the quasistatic  main
crack on the scale $\mathcal L$.
More precisely, the unstable onset of side branches, which we trigger
by modifying
dissipation locally, i.e.  within the process zone itself,
is proposed to
result from the competition between  the {\it growth} of secondary cracks
and their
{\it advection} in the displacement field of the main crack,
out of the near tip NLE zone where opening stresses remain
significant.

The article is organized as follows: in section \ref{sec:gelatin}, the physical
mechanisms of gelatin gel fracture are briefly
outlined, with emphasis put on the role of solvent viscosity.
The branching instability experiment is  described in section \ref{sec:experiment} and the
marginal nature of the branching onset is evidenced. This makes it
possible  in section \ref{sec:model} to propose a model featuring a single free
parameter, lumping together the NLE properties of the blunted
crack. Once this parameter is determined, the predictability of the
model is successfully tested. Possible generalization of the model to other
soft solids is discussed in part \ref{sec:discussion}.

\section{A reminder on gelatin gels}\label{sec:gelatin}

\subsection{Structure}
Gelatin, a biopolymer made of denatured collagen, can dissolve in
aqueous solvents (e.g. mixtures of water and glycerol) above the melting
temperature ($T_{m}\simeq 40^\circ$C). Upon
cooling below $T_{m}$, gelatin chains partially revert to
the triple helix conformation of the native
collagen, resulting in a network of rigid rods interspeded by
random coils \cite{Colby}. The mesh-size $\xi$ can be evaluated from the Young
elastic modulus $E$ assuming an entropic origin for the gel
elasticity:
$\xi\sim (k_{B}T/E)^{1/3}$. It is typically on the order of $10$
nm for Young  moduli in the $10$ kPa range.

The cross-linking triple helices are stabilized by weak physical H-bonds.
Accordingly, the gel
is thermoreversible. Moreover, it ages as revealed by the logarithmic
increase of $E$ with time, and exhibits slow stress relaxation
typical of soft glassy materials\cite{Rheo}.
However, on short times relevant to most fracture experiments, the polymer
network can sustain
shear. The relative motion of the solvent
with respect to the elastic network is a
collective,  diffusive process with a  coefficient
$D_{coll}\sim E\xi^{2}/\etab$  with $\etab$ the solvent viscosity.
Typically $D_{coll}\simeq 10^{-11}$ m$^{2}$.s$^{-1}$, an order of
magnitude which
makes solvent draining a very slow process on macroscopic length
scales so
that, for all practical purposes, the gel samples can be considered
quasi-incompressible.

\begin{figure}[h]
    \centering
    \includegraphics{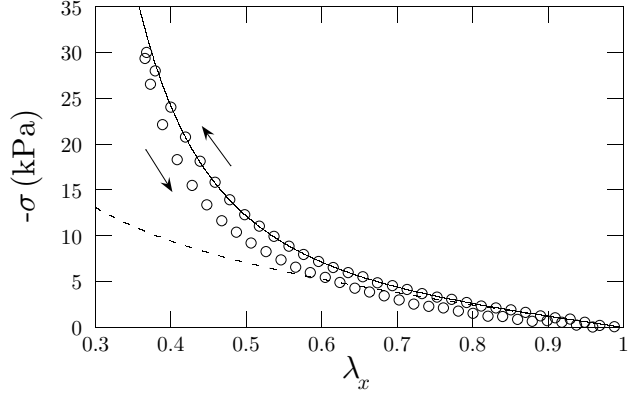}
    \caption{Non-linear elastic response of a cylinder of gelatin gel (5
    wt.\% in 60\%-40\% glycerol-water solvent)
    in uniaxial compression between well lubricated plates. $\sigma$ is the true (Cauchy) stress and
    $\lambda_{x}$ is the compression ratio. The solid curve is a fit
    using a strain energy density $W_{\mathrm{SH}}$ (see text) with
    $J_{m} = 2.3$.
    The initial slope ($\lambda_{x}\lesssim 1$) yields the Young modulus
    $E = 12.1$ kPa.  The dashed line
    corresponds to the neo-hookean elastic solid with the same small
    strain modulus. The small hysteresis between loading and
    unloading curves is probably indicative of slow stress
    relaxation\cite{Rheo} during the 6 s duration of the cycle.}
    \label{Fig:Load}
\end{figure}

Non-linear elasticity of gelatin gels exhibits strain hardening at
moderate strains. As shown on figure \ref{Fig:Load}, it does not obey the classical neo-hookean
constitutive law \cite{Treloar} (corresponding to a strain energy density functional
$W_{NH}(\lambda_{x},\lambda_{y},\lambda_{z})
= EJ_{1}/6$ with   $J_{1} =
\lambda_{x}^{2}+\lambda_{y}^{2}+\lambda_{z}^{2}-3$ an invariant
function of the principal stretch ratios, related by the
incompressibility condition $\lambda_{x}\lambda_{y}\lambda_{z}=1$).
The data obtained in uniaxial compression are, rather,
consistent with an empirical expression used in
numerical fracture studies as a simple model for severe strain hardening
\cite{HuiTip}:
$W_{\mathrm{SH}} = EJ_{m}(\exp(J_{1}/J_{m})-1)/6$ with $J_{m}\simeq
2.3$
suggesting that for elongations of order $\sqrt{J_{m}+3} \simeq$ 230\% the chains are already
stretched significantly taut.

\subsection{Fracture}\label{sec:gelfrac}

Experimentally, the rate-dependent fracture energy $\mathcal G(V)$
for a crack propagating steadily  at a markedly subsonic (or
``quasistatic") velocity $V$ reads~:

\begin{equation}
    \label{Eq:GV}
    \mathcal G = \mathcal G_{0}+\Gamma \etab V
    \end{equation}
with $\etab$ the  solvent viscosity, $\mathcal G_{0}\simeq 1$
J.m$^{-2}$ and $\Gamma \sim 10^{6}$. This linear relationship has
been interpreted \cite{Model} as resulting from the fact that in this physical
gel, fracture does  not proceed by chain scission but rather via unzipping
of the cross-links and subsequent pulling out of the overall chains
to the expense of the viscous drag  against the solvent.
$\mathcal G_{0}$ stems from the
plastic work for unzipping and, possibly, from the ``dehydration"
cost for exposing
the polymer chains to air. The rate-dependent term accounts for the
viscous losses against the solvent. $\Gamma$
 is predicted to scale as the squared  ratio of the
chain contour length ($\Lambda  \simeq 1 \,\mu$m) to the mesh size
of the network ($\xi \simeq 10$ nm).

Note that the stress level at the crack tip, given by $\mathcal
G/\Lambda$ is always several order of magnitude larger than the Young
modulus. This is precisely the case where elastic blunting of the
crack is predicted, hence where a marked NLE zone ahead of the crack
tip is expected \cite{Blunting}.

The sensitivity of the fracture energy to solvent viscosity makes
it possible to estimate the extension $d$ of the process zone where
energy is dissipated \cite{Model}. This is achieved by wetting the crack tip
opening with a drop of solvent of viscosity
$\etad<\etab$. Both drop and bulk solvents differ only by
their fractions  of glycerol in water which tend to equilibrate
via molecular diffusion, with a coefficient $D_{\mathrm{gly}}$. Experimentally, the
drop acts as a reservoir. The glycerol content
of the gel ahead of the moving tip adjusts from the drop
concentration to the bulk one
over the diffusive skin depth of order $D_{\mathrm{gly}}/V$.
At low enough velocities, the effective viscosity
\begin{equation}
    \etaf =
\Gamma^{-1} \frac{d\mathcal G}{dV}
    \label{Eq:etaeff}
\end{equation}
which accounts for the advance of ``rinsing"
within the process zone
is that of the drop liquid. As $V$ increases
above $V_{\mathrm{diff}}\simeq D_{\mathrm{gly}}/d$,
$\etaf (V)$  gradually grows up to $\etab$.
Measuring $V_{\mathrm{diff}}$ yields   $d \simeq 100$ nm.

This estimate was obtained originally \cite{Model} with a relatively small viscosity
contrast, $\etab/\etad \simeq 3$. Unexpectedly, on attempting to
reproduce the experiment with a larger contrast, a spectacular
branching instability was observed. It is the aim of this paper to
describe and interpret this phenomenon.

\section{Experimental}\label{sec:experiment}
\subsection{Material and methods}

Gel samples are composed  of $5$ wt.\% of gelatin (Sigma, ``300 Bloom" grade)
in aqueous solvents containing from
$\phi = 0$ to
$70$ wt.\% of glycerol in water.
The pregel solution is prepared by letting gelatin dissolve
in its solvent at $85^\circ$C under gentle stirring. It is poured
into rectangular molds (length $L = 300$ mm, width
$h = 30$ mm, and
thickness $e=10$ mm) and ``set" at $5^\circ$C for 10 hours.  Before
performing any experiment, the samples are left to equilibrate at
room temperature ($T\simeq 20^\circ$C) for 2 hours. Aging during the
duration of an experimental run (about 2 min) has been checked to be
negligible. The samples are characterized by their small strain Young
moduli $E$ (between $9$ and $15$ kPa, increasing with glycerol
content) and their solvent viscosity $\etab$  (from $1$ cP for
pure water to $22$ cP at $\phi = 70\%$).

\begin{figure}[h]
     \centering
    \includegraphics[scale = 0.75]{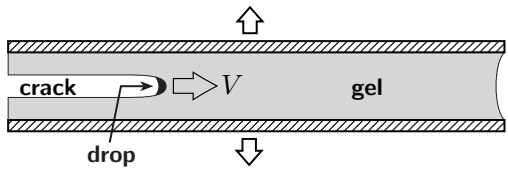}
    \caption{Schematic experimental setup. The gel sample is
    stretched between two rigid, parallel grips so that the energy release
    rate is imposed. The steady-state crack front velocity $V$ is measured.   }
    \label{Fig:Setup}
\end{figure}

Wetted crack experiments are performed with a setup fully described
elsewhere \cite{EPJE}. Gel plates  are stretched along their width by
imposing a
displacement $\Delta h$ to the rigid grips (see Fig.\ref{Fig:Setup}).
Due to the large aspect ratio $L/h=10$, uniform stretching is realized
in a significant portion of the sample,
thereby storing a prescribed amount of elastic energy $\mathcal G$
to be ultimately released per
unit area of a straight crack  propagating along the length.
$\mathcal G(\Delta h)$ is evaluated as $\mathcal W/(eL)$
with $\mathcal W$ the measured work required
to stretch
an un-notched sample by $\Delta h$. Note that, although the expression for $\mathcal G$ is
strictly valid for infinitely long samples \cite{Rivlin}, it is expected to provide
a reasonable approximation to within about $h/L=10^{-1}$.
The loading rate for the calibration test is chosen high enough for stress relaxation to be
negligible during the load (see eg. Fig \ref{Fig:Load}). Also the phenomenon reported
in the present study occurs on time scales short enough for internal
dissipation associated with remote stress relaxation  to be
negligible in comparison with the energy released by the fracture
itself.

Crack propagation is
initiated by cutting a notch   at one end of the  plate.
Away from the sample edges, when the
crack velocity reaches a steady value $V$ of order a few mm.s$^{-1}$,
a drop ($\simeq 200\,\mu$L)
of a water/glycerol
mixture of viscosity $\eta_{\mathrm{drop}}$ is quickly injected into the
tip opening where it remains trapped by capillarity and gravity (the
crack  travels  downward vertically).  The subsequent
dynamics of the crack tip is monitored at 10 frames  per second by a
video camera mounted on a traveling stage
so as to keep the crack front within the field of view.

\subsection{Characteristics of the branching instability}

\begin{figure}[h!]
    \centering
    \includegraphics[scale=1.0]{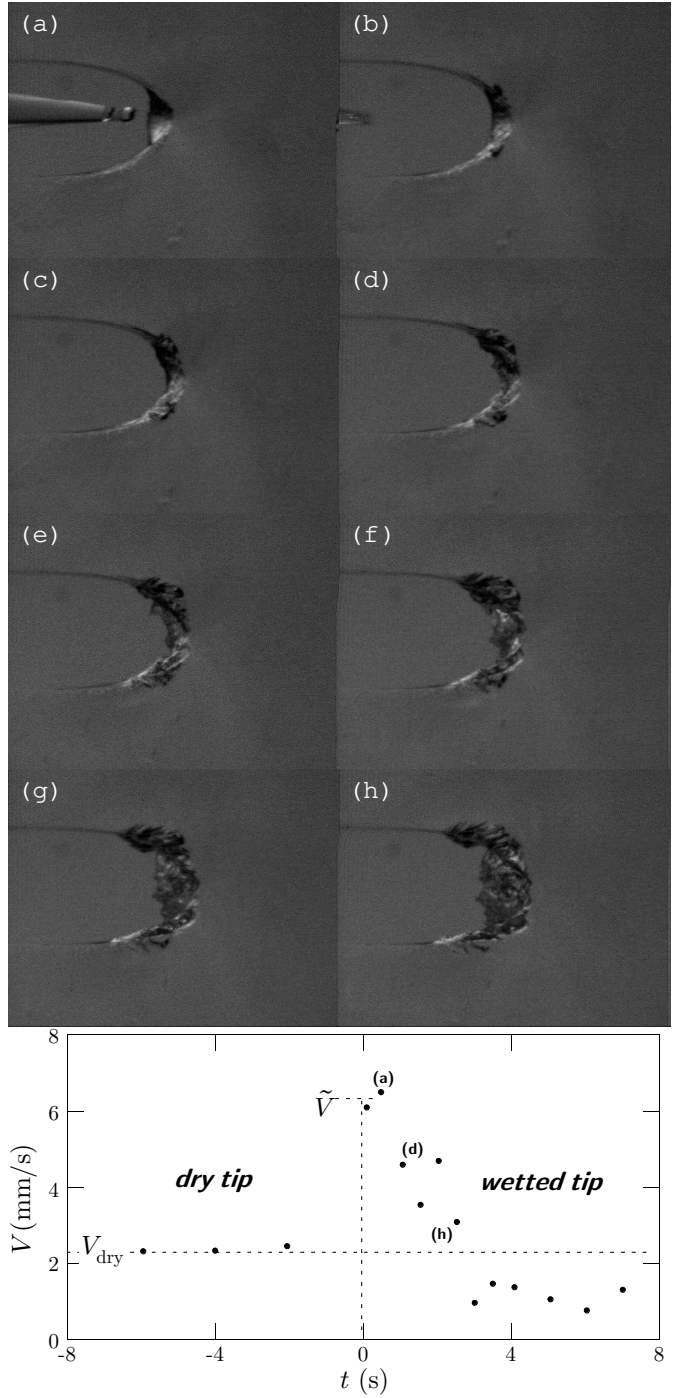}
   \caption{Snapshots ($\Delta t = 0.3 $ s)  extracted from a movie
    \cite{Video} showing the coarsening of a microcracked zone ahead of
    a crack  tip wetted at $t = 0$ by a drop of pure water in a $\phi =
    60\%$ gel. Crack tip opening is 12 mm ($\mathcal G = 65$
    J.m$^{-2}$). The graph displays the
    front velocity before and after tip wetting. }
    \label{Fig:Burst}
\end{figure}

Figure \ref{Fig:Burst} shows the first stage of the
instability triggered by
pure water ($\etad = 1$ cP) wetting the tip
of a gel with $\phi = 60$\% ($\etab =
11$  cP).  Within a few seconds, the gel ahead of the crack
is pervaded by a 3D damaged zone made of microcracks. The zone  coarsens
until the average front becomes flat. At that point, the
fracture process is almost inhibited (for the subsequent slow return to a
straight, dry crack, see movie \cite{Video}).

This spectacular, bursting response to a rather modest
environmental change is in marked contrast to our previous report
\cite{EPJE} on
the wetting by pure water of a $\phi = 30$\%  gel ($\eta_{\mathrm{bulk}} = 1.8$
cP). Then, while being markedly accelerated the crack remained
straight. Crack acceleration was ascribed to two  solvent effects:
firstly, wetting the tip prevents the extracted chains from being
exposed to air, hence a significant lowering of the threshold energy
$\mathcal G_{0}$; secondly, the water drop induces an osmotic imbalance, hence a glycerol
depletion ahead of the tip, as described in \ref{sec:gelfrac}. Both effects
contribute to cutting down
the fracture energy.  Since the energy release rate
is imposed, it results in the observed speeding up of the crack.
Fig.\ref{Fig:Burst} shows that, beyond this effect, local
solvent dilution can destabilize a crack tip as
well, depending
on the bulk viscosity.

\begin{figure}[h]
    \centering
    \includegraphics[scale = 1.0]{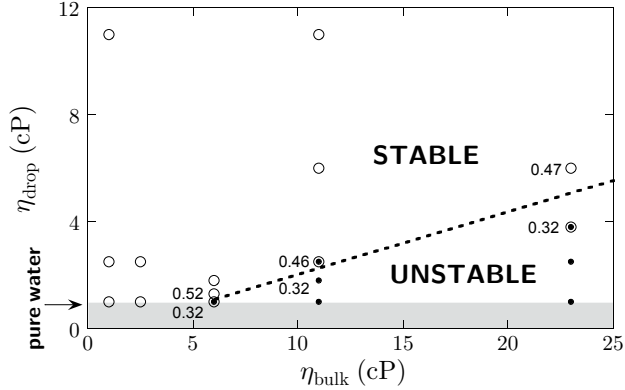}
    \caption{Stability diagram at constant crack tip opening (12 mm) for cracks in gels with bulk
    water/glycerol solvents
    of viscosities $\etab$ wetted by drops of viscosities
    $\etad$.  The shaded zone is out
    of experimental reach. Near critical data are
    labeled by the values of the compound parameter
    $(\eta_d/\etab).(\tilde V/V_{\mathrm{dry}})$ with $\tilde V$ and
    $V_{\mathrm{dry}}$ defined on Fig.\ref{Fig:Burst}. The  dashed  line is the
    {\it predicted} locus of critical cracks (see text, section \ref{sec:test}).}
    \label{Fig:Diagram}
\end{figure}

\begin{figure}[h]
    \centering
    \includegraphics[scale = 1.0]{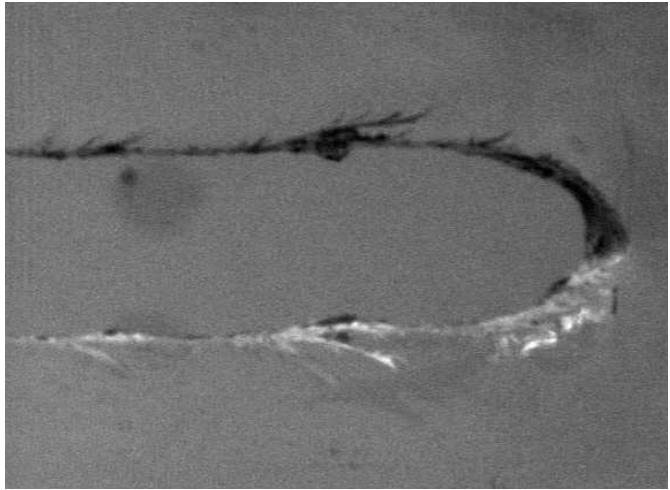}
    \caption{Wet crack in a critical state exhibiting aborted side branches.}
    \label{Fig:CriticalBranch}
\end{figure}

In order to clarify this point we have mapped the
stability of the crack onto the $(\etab, \etad)$ plane
for a prescribed $12$ mm crack tip opening. As displayed on Fig.\ref{Fig:Diagram},
branching occurs for  large enough
viscosity contrasts.
Since the tip solvent cannot be made more
inviscid than pure water, there is a minimal bulk viscosity,
corresponding to  $\phi = 50$\% below which the crack remains
stable whatever the tip environment.
Fig.\ref{Fig:CriticalBranch} shows  a snapshot
of a crack in this marginal  gel ($\phi = 50$\%), $4.8$ seconds after
being wetted by
a drop of pure water.
Though no thick damaged zone develops ahead of the tip, the crack
path exhibits distinct  undulations and aborted side branches.
Moreover, before these secondary cracks  stopped, they started themselves
to  branch, which we interpret as an
interrupted cascading process that, if unimpeded, would have led to
the microcracked zone of Fig.\ref{Fig:Burst}

Now, for a gel with a bulk viscosity above the marginal one, increasing the drop viscosity
from that of pure water changes the state of crack propagation from the unstable one shown on
Fig.\ref{Fig:Burst} to a stable one via a regime similar to that described above (see movies\cite{Video}).
It is therefore legitimate to term ``critical" this regime of
crack emission which marks systematically the frontier between
stable and unstable fracture.

\section{A schematic model}\label{sec:model}
Disregarding the complex, fully developed structure of Fig.\ref{Fig:Burst},
we now focus on the critical regime and propose a
criterion for the onset of instability that combines
the microscopic model of fracture energy specific for physical
hydrogels such as gelatin
and the more generic concept
of a  mesoscopic near-tip, non-linear elastic (NLE) zone.  As a starting point we
note that secondary cracks are emitted in the close vicinity
of the main tip. While they grow, they are progressively advected
away and gradually close on
approaching the
straight,  traction free crack edges.
Thus, there is a competition between growth and advection rates. We
propose that only those cracks which manage to grow significantly
before being advected out of the near-tip zone, where opening stresses
are significant, can in turn branch, hence trigger a cascading
process.   We therefore describe branching as a {\it convective}
instability\footnote{Surface tension could help nucleating cracks at the pinned contact line of the
drop, as described for very soft gels in reference \cite{Behringer}. This is not observed here, however.}.

\subsection{Near-tip stress field of a blunted crack}

Branching requires nucleation and growth of secondary cracks which
eventually compete with the initial main crack to release
elastic energy. We   assume in the following that nucleation sites, or flaws,
are provided by the inhomogeneities --- of either
toughness or stiffness --- of the randomly crosslinked polymer network.

As a consequence of elastic blunting, associated to strain-hardening
of the gel, the region where
opening stresses are high is along the crack face rather than
directly ahead of the crack tip, a qualitative feature which is
clearly revealed by finite
element simulations \cite{HuiTip}.  This suggests
that branching probably originates from nucleation and growth of edge cracks in this
quasi-uniaxially stressed zone. Its  extension   is
provided, in a first approximation, by the radius of curvature of the
near tip parabolic crack opening.
This  is given, within a multiplicative constant
of order unity, by $\mathcal G/E$. This classical result of LEFM
\cite{Lawn}  remains  valid in the NLE case, at least for
a neo-hookean solid (see e.g. \cite{HuiTip}), and is certainly dimensionally
correct for
more strongly strain hardening materials.

\begin{figure}[h]
    \centering
    \includegraphics[scale=0.75]{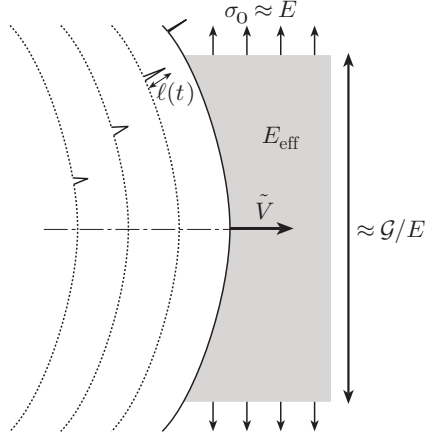}
    \caption{Schematic representation of a secondary crack growth and
    advection in the NLE zone of a blunted main crack. }
    \label{Fig:Schematic}
\end{figure}

We now aim at writing a growth
equation for an edge crack in the NLE stress field. This is a
desperately complex mechanical problem. We therefore make one
further step and, based on the previously described structure of the
near-tip stress field \cite{HuiTip}, we schematize the blunted elastic zone as a
strip of size $\sim\mathcal G/E$, uniformally stretched by a stress
$\sigma_{0}$ perpendicular to the fracture plane (see Fig.\ref{Fig:Schematic}).
The material is assumed to have strain-hardened
up to a state characterized by a uniform small strain modulus $E_{\mathrm{eff}} >E$.

Progressive building-up of strain hardening on approaching the
blunted crack tip is now replaced by a stepwise jump in small
strain modulus at a distance $\mathcal G/E$. The unknown stress
$\sigma_{0}$ is determined according to the compatibility requirement that LEFM
should hold in the outer region. There, the stress intensity factor
\cite{Lawn}
$K_{I} \sim \sqrt{\mathcal G E}$, hence the stress at a distance
$\mathcal G /E$ is $\sigma_{0} \sim K_{I}/\sqrt{\mathcal G/E} \sim E$.
Note that this rough schematization of the blunted elastic zone is akin to that
proposed in \cite{Blunting} on the basis of slightly different
arguments.

\subsection{Growth mechanism for a secondary crack}

Since we are interested in the onset of unstable branches, we
assume  the length $\ell$ of the secondary crack to  remain
small as compared to the extension $\mathcal G/E$ of the blunted crack zone, so that
it can be treated as an edge crack which does not interfere
significantly with the stress field of the main crack.
For the sake of simplicity,
we estimate the corresponding energy release rate
$\mathcal G_{\mathrm{branch}}$ according to  LEFM \cite{Lawn}
in a material with an effective Young modulus
$E_{\mathrm{eff}}$. Within a factor of order unity:
$\mathcal G_{\mathrm{branch}} = (\pi\sigma_{0}^{2}/E_{\mathrm{eff}})\ell
= \beta E\ell$, where we have introduced a
dimensionless factor $\beta \approx E/E_{\mathrm{eff}}$ which is indicative of
the level of strain-hardening.
The clear separation of lengthscales, $d\ll\mathcal G/E$, makes it legitimate to
 assume that the edge crack dynamics on scale $\mathcal G/E$ is ruled by the same
physical mechanism as the main crack. The growth equation is thus
obtained by equating $\mathcal G_{\mathrm{branch}}$  with the fracture energy
$\mathcal G $ given by (\ref{Eq:GV}), replacing $V$ by $d\ell /dt$:

\begin{equation}
    \beta E \ell = \mathcal G_{0} + \Gamma\eta_{\mathrm{branch}} \frac{d\ell}{dt}
    \label{Eq:Growth}
\end{equation}
where $\eta_{\mathrm{branch}}$ is the effective solvent viscosity ``felt"
by the chains pulled out of the process zone   ahead of the secondary
crack.

As discussed in the following section,  a possible rate-dependence of
$\eta_{\mathrm{branch}}$ can be ruled out.
Solving eq. (\ref{Eq:Rate}) for $\ell(t)$ is therefore
straightforward~: $\ell(t) = \ell_{c} +(\ell_{c}-\ell_{0})\exp
(t/\tau)$. Accordingly,
any  supercritical seed crack of initial size $\ell_{0}> \ell_{c}=\mathcal
G_{0}/\beta E$ will
grow exponentially over a characteristic time
\begin{equation}
    \tau =
\frac{\Gamma}{\beta}
\frac{\eta_{\mathrm{branch}}}{E}
    \label{Eq:Rate}
\end{equation}

If the secondary crack is left to grow in the constant stress field it will eventually,
at times on the order of $\tau$, compete
with the main crack for releasing energy, and give rise, in turn,
to side branches. However, while the main crack grows, the root of the secondary crack
is advected away from the main tip, eventually leaving the NLE zone where sufficient stress
exists to permit branch growth. The stability criterion discussed in
the following is based on the competition between these two effects.

\subsection{A criterion for stable, wetted cracks}

The life time of a secondary crack is given by the duration of its
travel along the blunted tip zone as it is advected away from the apex of
the main crack. An upper value for this duration is the time
required for crossing a region of size $\mathcal G/E$ at $\tilde V$, the
velocity of the main crack tip in its wetted state~:
\begin{equation}
    T_{\mathrm{adv}} = \frac{1}{\tilde V} \frac{\mathcal G}{E}
    \label{Eq:Tadv}
\end{equation}

As long as $\tau > T_{\mathrm{adv}}$,  all secondary cracks will stay in
their embryonic states and never compete with the main crack which
will therefore remain stable.
Since, according to (\ref{Eq:etaeff}), $\tilde V$ is given implicitly
for a prescribed $\mathcal G$ by

\begin{equation}
    \mathcal G = \mathcal G_{0}
+\Gamma\etaf(\tilde V)\tilde V
    \label{Eq:Vtilde}
\end{equation}
with $\mathcal G_{0}$ being usually negligible (see
Fig.\ref{Fig:Eta}), the stability criterion reads: $\eta_{\mathrm{branch}}/\etaf(\tilde V)>\beta$.

At this stage, it is crucial to realize that the dynamics of the
secondary crack is ruled by the effective viscosity $\eta_{\mathrm{branch}}$ which depends  on glycerol diffusion, i.e. on
the velocity of the branch tip with respect
to the liquid drop. Since the latter is driven by capillarity and
gravity along with the main crack, the relevant velocity is  that
at which the secondary tip cuts into the primary crack lips, i.e. $d\ell
/dt$. We have therefore $\eta_{\mathrm{branch}} = \etaf(d\ell/dt)$.
Since $d\ell/dt$  remains much smaller than the velocity $\tilde V$ of the
main tip during the maturation phase (secondary crack growth times
$t<\tau$), one can expect that
$\etaf(d\ell/dt)<\etaf(\tilde V)$.
In fact, it can be shown that $\eta_{\mathrm{branch}} = \etad$
(see appendix)  so that, finally, the criterion for a
crack to remain {\it stable} at a quasistatic velocity $\tilde V$ reads
simply:
\begin{equation}
   \frac{\etad}{\etaf (\tilde V)}> \beta
    \label{Eq:criterion}
\end{equation}

\subsection{Experimental tests  of the model}\label{sec:test}

The above stability criterion (\ref{Eq:criterion}), is expressed in
terms of  the effective viscosity functional, defined by
(\ref{Eq:etaeff}), and   parameter $\beta$. The
former accounts for the local modification of the microscopic process zone via
solute diffusion, the latter is aimed at catching the NLE
modification of the gel at the mesoscopic  scale of the blunted crack
opening.

Let us note in the first place that, in the destabilizing
configuration where the drop
is less viscid than the bulk solvent, the inequality
$\tilde V$:  $\etad/\etaf(\tilde V) \leq 1$
holds
for any combinations of $\tilde V$, $\etad$ and $\etab$, including
those corresponding to stable cracks.
Hence, according to (\ref{Eq:criterion}), $\beta <1$. So, in spite
of the numerous multiplicative constants of order unity hidden
in $\beta$, this parameter therefore retains its  physical
flavour as a qualitative indicator of the level of strain hardening~: $\beta\sim E/E_{\mathrm{eff}}<1$.

\begin{figure}[h]
    \centering
    \includegraphics[scale=1.0]{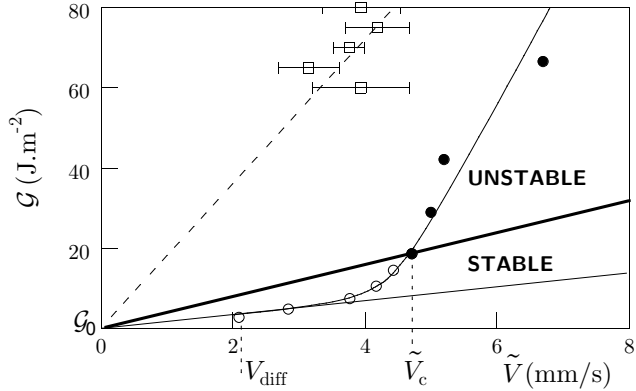}
    \caption{Stability diagram  in the fracture energy  {\it vs.} crack-tip
    velocity plane.
    Unwetted ($\Box$) cracks; stable ($\bigcirc$)  and unstable ($\bullet$) cracks
    wetted by pure water
    (for unstable cracks, $\tilde V$ is the front velocity
    measured just after wetting, see Fig.\ref{Fig:Burst}).
    The curve is a guide for the eyes.
    Slopes of the thin, thick and dashed lines are in the ratio
    $\eta_{\mathrm{tip}}:\eta_{\mathrm{eff}}(\tilde V_{c}):\eta_{\mathrm{bulk}}$. $V_{\mathrm{diff}}$ and
    $\tilde V_{c}$ are defined in the text.}
    \label{Fig:Eta}
\end{figure}

Furthermore, since $\etaf$ is an increasing function of the tip velocity,
(\ref{Eq:criterion}) predicts that cracks are stable at low
enough velocities. This is a counter intuitive result, owing to
the proposed convective nature of the instability,  since for a slow main
crack, side  branches are only slowly advected away from the tip
zone, hence  have  {\it a priori} more time to grow.
The fact that this direct effect of $\tilde V$ disappears in (\ref{Eq:criterion}) can be
traced back to the
$V$-dependence of the NLE length scale $\mathcal G(V)/E$.
A slow crack offers a  shorter NLE
zone for branches to grow. The existence of a lower critical velocity for a
given pair $\etad <\etab$ is therefore a strong test of the validity
of our schematic model.
We have checked it by increasing stepwise  the crack
opening, hence $\mathcal G$, while keeping the tip wet by continuous
solvent dripping. Fig.\ref{Fig:Eta} shows the results for a pure
water drop wetting a $\phi = 60\%$ gel.
Indeed, the crack remains stable as long as $\tilde
V < \tilde V_{c}$ with $\tilde V_{c}\simeq 4.7$ mm.s$^{-1}$.

This enables us to evaluate $\beta$ quantitatively.
Following the above discussion,  for $\tilde
V<V_{\mathrm{diff}}$ the tip is fully rinsed and $\etaf \simeq
\etad$, hence $\mathcal G\sim
\etad\Gamma\tilde V$. Fitting the low velocity data accordingly yields
$\Gamma \simeq 1.6\times 10^{6}$, a value compatible with that of non-wetted
cracks in a companion sample of the same gel  (see Fig.
\ref{Fig:Eta}).
At the critical point, we measure
$\etaf(\tilde V_{c}) = (\mathcal G(\tilde V_{c})-\mathcal
G_{0})/(\Gamma\tilde V_{c})$ from which we deduce that $\beta\simeq
0.42$.

Finally, we  return to the stability diagram of Fig.\ref{Fig:Diagram}.
Though chosen for  the sake of its experimental simplicity, the constant crack tip
opening prescription  is theoretically awkward  since it corresponds
to  a non trivial section of the
critical surface in the 3D control parameter space
($\eta_{\mathrm{bulk}},\eta_{\mathrm{drop}}, \tilde V$).
It is nevertheless possible to make use of this diagram for an independent  check of the
model by  recasting criterion (\ref{Eq:criterion}) in terms of $V_{\mathrm{dry}}$ (resp. $\tilde V$), the
crack velocities before (resp. just after) tip wetting (see
Fig.\ref{Fig:Burst}). Neglecting
$\mathcal G_{0}$, the constant opening imposes
$\mathcal G\simeq \eta_{\mathrm{bulk}}\Gamma V_{\mathrm{dry}} \simeq \eta_{\mathrm{eff}}(\tilde
V)\Gamma \tilde
V$, hence the stability condition~:
$(\eta_{\mathrm{drop}}/\eta_{\mathrm{bulk}}).(\tilde V/V_{\mathrm{dry}}) > \beta$.
We have computed the left hand side compound parameter for near
critical data on both sides of the
bifurcation (see Fig.\ref{Fig:Diagram}). Remarkably, the figures systematically bracket the value
$\beta = 0.42$ determined from the critical velocity.

\section{Concluding remarks}\label{sec:discussion}

We have identified and analyzed a new branching instability occurring in
the tip vicinity of a quasistatic crack in gelatin hydrogels.
We have  proposed what we think
is the first predictive model of a
{\it convective} crack
branching instability in a disordered elastic material.
The key ingredient  is the existence of a mesoscopic zone
ahead of the crack where  residual opening stresses are large enough
for promoting side branching.  The finite size of this near tip zone,
which scales with $\mathcal G/E$, results in a finite advection time of
 incipient branches away from the main crack. Whether catastrophic
crack growth occurs during this time chiefly depends on the
rate-dependence of the
fracture energy. In the case of gelatin, it
can be conveniently tailored by a local ``osmotic" control of  the process zone.

Beyond the deciphering of a puzzling phenomenon which might, at
first sight, appear as an idiosyncrasy of gelatin gels, we think
that our analysis
lends support  to the emerging idea that the non-linear elastic field
which bridges between the process zone and the generic linear elastic
region, plays a crucial role in crack dynamics and, more
specifically, in crack front instabilities \cite{Bouchbinder}.
Though branching instabilities have been mostly documented for fast
cracks,
our study indicates that they could prevail as well for slow ones
over wide domains of their
mechanical control parameters (crack tip velocity, tip
environment, structural disorder,\ldots).

To go one step further along this line, let us analyze a case  where
quasistatic crack propagation would be ruled by a fracture energy
that depends as a sublinear power-law on the crack velocity, say:

$$\mathcal G = \mathcal G_{0}[1+(V/V_{0})^\alpha ]\,\,\mathrm
{with}\,\, \alpha <1$$
Such a functional form is typical of elastomers \cite{Gent} and has
been reported for gels made of solvated triblock copolymers as well
\cite{Triblock}. In the case of elastomers, the material-dependent reference velocity
$V_{0}$ depends on temperature, presumably via the visco-elastic
spectrum of the material. We assume that the generalized small-scale
yielding approximation ($d\ll\mathcal G/E\ll L$) holds here.

The growth equation $\mathcal G(\dot \ell) = \beta E\ell$ for an edge
crack nucleated within the primary blunted tip
region can be solved analytically. It is straightforward to establish
that the degree of supercriticality of the crack, introduced for the
sake of simplicity as
$\epsilon(t) = \ell(t)/\ell_{c}-1$ with $\ell_{c} = \mathcal
G_{0}/\beta E$, reads:
$$\epsilon(t) = \frac{\epsilon_{0}}{(1-t/\tau')^{1/\gamma}}
\,\,\mathrm{with}\,\,\tau'  =
\frac{\ell_{c}}{V_{0}\gamma\epsilon_{0}^\gamma} \,\,\mathrm{and}
\,\,\gamma = \alpha^{-1}-1>0$$
Crack length divergence at finite time $\tau'$ signals an
instability. It is important to note, however, that the time $\tau'$ for
catastrophic failure depends on $\epsilon_{0}=\epsilon(0)$, the initial degree of
supercriticality
of the unstable crack seed.

Due to advection of the material in the tip region, the main crack will however remain stable
against side branching provided that $V\tau'
>   \mathcal G(V)/E$, i.e.

\begin{equation}
    \frac{V/V_{0}}{1+(V/V_{0})^\alpha }>
\beta\gamma\epsilon_{0}^\gamma
    \label{eq:viscocriterion}
\end{equation}

Here, in contradistinction with the $\alpha =1$ case, a  secondary crack will close back
before exploding if the
main one is driven at a  {\it high} enough velocity (the rhs of
(\ref{eq:viscocriterion}) is an increasing function of $V/V_{0}$). Yet, this upper
critical velocity depends on the strength $\epsilon_{0}$ of the nucleation seed.
The larger the flaw, the more
destabilizing it is for the main crack.
Such an instability would possibly manifest itself as follows: in the absence
of any external triggering,  a  crack
running steadily at a velocity $V$ would eventually become unstable,
provided it meets a flaw of sufficient strength. Since, according to
(\ref{eq:viscocriterion}), the size of the critical seed increases
with $V$, the higher the
velocity, the lower the probability of meeting such a nucleation site
and the longer, statistically, the crack would remain stable.

This example shows  that, even in its simplest version, our
model predicts a wealth of qualitative behaviors. It prompts us to
have a fresh look at the complex, intermittent dynamics reported in the case of
triblock copolymer gels \cite{Triblock}, and still unexplained. It also points to the
important role played by the structural disorder on branching and suggests to study
controlled, inhomogeneous materials such as filled elastomers or
bicomponent, phase-separated polymer hydrogels.

\section*{Appendix}

In this appendix, we show that the velocity of a branch is, in its maturing phase,
small enough so that its process zone is fully rinsed by the drop i.e. that
$\eta_{\mathrm{branch}} = \etad$ holds in eq. \ref{Eq:Rate}.

Consider a supercritical crack seed of initial length $\ell(0) =
(1+\epsilon_{0})\ell_{c}$. For $t<\tau$, the crack velocity remains close
to its initial value $v_{0} = \epsilon_{0}\ell_{c}/\tau = \epsilon_{0}\mathcal
G_{0}/(\eta_{\mathrm{branch}}\Gamma)$. Since $\eta_{\mathrm{branch}} \geq \etad \geq \eta_{\mathrm{water}}$,
and $\mathcal G_{0}\simeq 1$ J.m$^{-2}$, $v_0 < \epsilon_{0}\times 1$ mm.s$^{-1}$.
We argue in the following that, in our experiments, $\epsilon_{0}$ remains small enough so
that $v_0 < V_{\mathrm{diff}}$.

Our model assumes that  ``flaws" preexist is the sample and
act as  edge crack seeds when reached
by the blunted tip zone
of the main crack. As already mentioned, flaws are ascribed to frozen
fluctuations of the polymer network structure \cite{Frozen},  e.g. via its crosslink density.
Such inhomogeneities remains after coarse-graining up to the scale $d$
of the process zone and result in a spatially  modulated  gel
strength about an average value
$\bar{\mathcal G}_{0}$, as evidenced by the  roughness of fracture
surfaces in gelatin gels \cite{Morpho}. An estimate of the maximum amplitude of
$\mathcal G_{0}$ variations, $\Delta \mathcal G_{0}/\bar{\mathcal G}_{0}\simeq
14\%$, has been obtained by analyzing the
pinning of very slow crack fronts induced by ``tough" spots
\cite{Morpho}.

According to Griffith's theory \cite{Lawn}, crack initiation results
from a balance
between the fracture energy cost $\mathcal G_{0}$ and the elastic
energy released in a deformed region scaling with $\ell$. That is
a nucleation process, with an activation energy
corresponding to a critical crack length $\ell_{c}$.
With $\bar{\mathcal G}_{0}\simeq 1$ J.m$^{-2}$, $\ell_{c}=\bar{\mathcal
G}_{0}/\beta E\simeq
100\,\mu$m. The activation barrier $\mathcal E_{\mathrm{act}}\sim \bar{\mathcal
G_{0}}\ell_{c}e$, with $e$ the lateral extension of the
crack,  is always orders of magnitude larger than $k_{B}T$ whence
thermal activation cannot be responsible for the distribution of
initial crack lengths.
We propose here that it rather reflects the statistical
distribution of the flaw strengths  $\mathcal G_{0}$, at the scale of
$d$ and therefore
identify the supercriticality index $\epsilon_{0}$ with $\Delta
\mathcal G_{0}/\bar{\mathcal G}_{0}$ so that
$v_{0}<1$ mm.s$^{-1}$. With $D_{\mathrm{gly}}(\phi = 60\% )=2\times 10^{-10}$
m$^{2}$.s$^{-1}$ and $d\simeq 100$ nm, $V_{\mathrm{diff}}\simeq 2$ mm.s$^{-1}$.
Since $v_{0}<V_{\mathrm{diff}}$,
one may therefore safely consider that $\eta_{\mathrm{branch}} \simeq
\etaf(v_{0}) \simeq \etad$.

\acknowledgements{
We thank Christiane Caroli for constructive criticism and careful
reading of the manuscript, and David Martina for his serendipitous
discovery of the instability.}

\end{document}